\documentclass[aps,twocolumn,showpacs,floats]{revtex4}
\usepackage{graphicx}

\begin{document}

\title{ Binding energy of a holographic deuteron and tritium in anti-de-Sitter
space/conformal field theory (AdS/CFT)
 }

\author{M.\ R.\ Pahlavani
\footnote{m.pahlavani@umz.ac.ir}
       }
\author{J.\ Sadeghi
\footnote{pouriya@ipm.ir}
       }
\author{R.\ Morad
\footnote{r.morad@umz.ac.ir}
       }
\affiliation{
Department of Physics, Faculty of Basic science,Mazandaran University, P.O.Box 47416-1467, Babolsar, Iran\\
             }
\begin{abstract}
In the large 't Hooft coupling limit, the hadronic size of baryon is
small and nucleon-nucleon potential is obtained from massless
pseudo-scalar exchanges and an infinite tower of spin one mesons
exchanges. In this paper we use the holographic nucleon-nucleon
interaction and obtain the corresponding potential and binding
energy for deuteron and tritium nuclei. The obtained potentials are
repulsive at short distances and clearly become zero by increasing
distance as we expected.

\end{abstract}


\maketitle
\nonstopmode
\section{Introduction}
One application of AdS/CFT duality is in low energy hadron dynamics
[1] that is referred as holographic QCD or AdS/QCD. This method
express two related issues from opposite directions, one from string
theory [2,3] and the other from low energy chiral effective field
theory of mesons and baryons[4,5]. From the string theory viewpoint,
what extent the gravity theory in the bulk sector in a controlled
weak coupling limit, is interested that is called, via duality, the
strongly coupled dynamics of QCD. On the other hand, from the low
energy chiral effective field theory outlook the purpose is whether
holographic QCD can make clear predictions on processes which are
difficult to study by QCD proper.
\\There are many remarkable of this type examples [6-11] such as
chiral dynamics of hadrons, in particular baryons at low energy that
are typically of strong-coupling QCD and very difficult to obtain by
QCD techniques.
\\Between the holographic models of QCD suggested recently, the Sakai
and Sugimoto (SS) model [2] is one of the most interesting and
realistic model, because of accurate results of this model. For
example the predicted results from $SS$ model on glueball spectrum
of pure QCD are in a good agreement with lattice simulation[12].
Also this model successfully described baryons and their
interactions with mesons[2,13,14]. This is a $D_4-D_8$ model that
involves a large number of colors, $N_c$ large 't Hooft coupling
$\lambda$ and quenching of fermions.
\\ The holographic baryon in the $D_4-D_8$ model is same as the
skyrmion of chiral perturbation theory. Actually 'holographic
baryon' is a direct uplift of skyrmion in the holographic picture,
but considering baryons as solitons is not a proper way to obtain
the nucleon-nucleon interaction.It is arises because finding a
suitable configuration is impossible for such complicated solitons.
\\In order to study baryons interaction in large distances, where the
inter-baryon distance is large compared with the size of the
baryons, it is a good approximation to consider baryons as a
point-like particles. In this case, the interactions can be all
ascribed by exchange of light particles such as mesons and one can
find the baryon-baryon interaction with the Feynman diagrams using
cubic interaction vertices including baryon currents and light
mesons [15].
\\ Fortunately from the $D_4-D_8$ holographic QCD model, all
nucleon-meson coupling constants, at least in large $\lambda N_c$,
is obtained [13]. Also some of these coupling constants such as the
axial coupling to pions ,$g_A$ and vector meson couplings $g_{\rho
\mathcal{NN}}$ and $g_{\omega \mathcal{NN}}$ are good agreed with
experimental data. Recently the nucleon-nucleon potential in
holographic picture, is studied using the meson exchange method
[16].
\\ In this paper we are going to calculate the binding energy of
light nuclei such as deuteron and tritium nuclei using the
$AdS/QCD$. So we use the $D_4-D_8$ model, nucleon-meson interaction
and nucleon-nucleon potential. Then we consider the exchange of
pions, isospin singlet vector mesons, isospin triplet vector mesons
and triplet axial-vector mesons in this potential. The minimum of
the nuclei potential is considered as nuclei binding energy. Finally
we obtain the radius of these nuclei in the holographic picture.

\section{A $\mathbf{D_4-D_8}$ Holographic QCD}
In this model $N_c$ stack of $D_4-$branes and $N_f D_8-$branes are
considered in the background of
Type\uppercase\expandafter{\romannumeral 2}\,A \,superstring [2] and
the flavor symmetries of the quark sector are embedded into a
$U(N_f)$ gauge symmetry in $R^{1+3}\times I$. By restricting to the
modes that are localized near the origin of the fifth direction,
which is topologically an interval, we can arrive to the four
dimensional low energy physics. Also the massless part of the theory
is pure $U(N_c)$ Yang-Mills theory because the fermions are given
anti-periodic boundary condition.
\\ In the large $N_c$ limit, the $D_4$ branes dynamics is dual to a
closed string theory in the curved background with flux in
accordance with general AdS/CFT idea.
 In the large 't Hooft coupling limit $(\lambda=g_{YM}^2 N_c >>1)$
and neglecting the gravitational backreaction from the $D_8$ branes
the metric is [17]

\begin{eqnarray}
ds^2
&=&\left(\frac{U}{R}\right)^{3/2}\left(\eta_{\mu\nu}dx^{\mu}dx^{\nu}+f(U)d\tau^2\right)
\nonumber \\
&+&
\left(\frac{R}{U}\right)^{3/2}\left(\frac{dU^2}{f(U)}+U^2d\Omega_4^2\right),
\label{1}
\end{eqnarray}

where $R^3=\pi g_sN_cl_s^3$ and $f(U)=1-U_{KK}^3/U^3$. The
coordinate $\tau$ is compactified as $\tau=\tau+\delta\tau$ with
$\delta\tau=4\pi R^{3/2}/(3U_{KK}^{1/2})$.
\\ The effective action on a $D_8$ brane embedding in $D_4$
background has the following form

\begin{eqnarray}
S_{D8}=&-& \mu_8 \int {\rm
d}^9x\,e^{-\phi}\sqrt{-\det\left(g_{MN}+2\pi\alpha^{\prime}F_{MN}\right)}
\nonumber \\
&+&\mu_8\int\, C_{3} \wedge \, Tr e^{2\pi\alpha' F }, \label{2}
\end{eqnarray}

with $ \mu_8=\frac{2\pi}{(2\pi l_s)^{9}} $. By introducing the
conformal coordinate $w$ instead of holographic coordinate $U$ as

\begin{eqnarray}
w&=&\int_{U_{KK}}^U\frac{R^{3/2}dU^\prime}{\sqrt{{U^\prime}^3-U_{KK}^3}} \:,\label{3}
\end{eqnarray}

the noncompact 5D part of the metric is conformally flat, then the
induced metric on $D_8$ brane has the following form

\begin{equation}
g_{8+1}=\frac{U^{3/2}(w)}{R^{3/2}}\left(dw^2+\eta_{\mu\nu}dx^{\mu}dx^{\nu}\right)
+\frac{R^{3/2}}{U^{1/2}(w)} d\Omega_4^2\:, \label{4}
\end{equation}

where $U_{KK}=2/9 g_{YM}^2N_cM_{KK}l_s^2$. $M_{KK}$, $\lambda$ and
$N_c$ determine all the physical scales such as the QCD scale and
the pion decay constant.
\\In the low energy limit, the worldvolume dynamics of the
multi-$D_8$ brane system gives the following Yang-Mills action with
a Chern-Simons term as

\begin{eqnarray}
& & \frac14\;\int_{4+1}\sqrt{-g_{4+1}}
\;\frac{e^{-\Phi}V_{S^4}}{2\pi
(2\pi l_s)^5} \;\rm tr F_{\hat m \hat n}F^{\hat m\hat n}\nonumber\\
& &+\frac{N_c}{24\pi^2}\int_{4+1}\omega_{5}({\cal A})\: \:
,\label{5}
\end{eqnarray}

where $V_{S^4}$ is the $S^4$ volume while the dilaton is

\begin{eqnarray}
e^{-\Phi}=\frac{1}{g_s}\left(\frac{R}{U}\right)^{3/4} \: , \label{6}
\end{eqnarray}

and $d\omega_5(A)=\rm tr F^3$.

\section{Baryon Holography and Nucleon-Nucleon Potential}

Witten introduced a $D_4$ brane wrapping the compact $S^4$ as a
baryon vertex on the 5D space-time [18]. It is shown that a $D_4$
brane wrapping $S^4$ looks like an object with electric charge with
respect to the gauge field on $D_8$ and it is possible to say that
$D_4$ brane spread inside $D_8$ brane as an instanton. The size of
this instanton is determined by minimizing its total energy [13,14],
which is combined mass and coulomb energy,

\begin{equation}
\rho_{baryon}\sim \frac{9.6}{M_{KK}\sqrt{g_{YM}^2N}} \:.\label{7}
\end{equation}

Thus, in the large 't Hooft coupling limit, instantonic baryon is a
small object in 5 dimension and baryon can be considered as a
point-like quantum field in 5D. In consequence, there should
couplings between this quantum field and the 5 dimensional gauge
field moreover the standard Dirac kinetic and a position dependent
mass term [19].
\\The action involving
the baryon field and the gauge field in the conformal coordinate
$(x^{\mu},w)$ is written as
\begin{eqnarray}
\int d^4 x dw &&[-i\bar{\cal B}\gamma^m D_m {\cal B} -i
m_b(w)\bar{\cal B}{\cal B}+
\nonumber \\
&&g_5(w){\rho_{baryon}^2\over e^2(w)}\bar{\cal
B}\gamma^{mn}F_{mn}{\cal B} ]-\nonumber \\ \int d^4x dw &&{1\over 4
e^2(w)}\rm tr F_{mn}F^{mn}. \, \, \, \, \,\: \label{8}
\end{eqnarray}
$g_5(w)$ is an unknown function of $w$ that only evaluated at $w=0$, $g_5(0)=2\pi^2/3.$
\\Since the four-dimensional low energy physics is found by
restricting to the modes that are localized near the origin of fifth
direction $ w$, the physical 4D nucleons would arise as the lowest
eigenmodes of the 5D baryon along $w$ coordinate. Thus the
five-dimensional action, equation (8), must to be reduced to four
dimensions. It can be done by applying the mode expansion for the
baryon field and the gauge field and plugging these to the baryon
action.
\\On one hand, the gauge field $A_\mu$, in the $A_5=0$ gauge, has
the following mode expansion
\begin{equation}
A_\mu(x,w)=i\alpha_\mu(x)\psi_0(w) +i\beta_\mu(x)+\sum_n
a_\mu^{(n)}(x)\psi_{(n)}(w)\:.\label{9}
\end{equation}
The eigenmode analysis was done by Sakai and Sugimoto in [2]
previously. We only note that $\psi_{(2k+1)}(w)$ is even, while
$\psi_{(2k)}(w)$ is odd under $w\rightarrow -w$, corresponding to
vector and axial-vector mesons respectively. Also the eigenfunctions
$\psi_{(n)}$ obey the following equation according to [2]
\begin{equation}
-K^{-1/3}\partial_{\omega} (K^{1/3} \partial_{\omega} \psi_{(n)} ) =(U_{KK}^2 M_{KK}^2)\lambda_n \psi_{(n)}
 .\label{10}
\end{equation}
where $K=(\frac{U}{U_{KK}})^3$. Also they satisfy the orthonormality
condition
\begin{equation}
\int dw\,\frac{e^{-\Phi}V_{S^4}}{4\pi (2\pi
l_s)^5}\,\psi_{(n)}(w)^*\psi_{(m)}(w)=\delta_{nm}\:.\label{11}
\end{equation}
We have to solve (10) with the normalization condition given by (11)
to find the eigenfunction $\psi_{(n)}$. These equations solved
numerically by means of a shooting method. The corresponding
computations are given in [2] in details.
\\On the other hand, the nucleon field can be expanded as ${\cal
B}_{L,R}(x^\mu,w)=B_{L,R}(x^\mu)f_{L,R}(w)$ where $\gamma^5
B_{L,R}=\pm B_{L,R}$ are 4D chiral components. $f_{L,R}(w)$ are
profile functions that satisfies the following conditions in the
interval $w\in[-w_{max},w_{max}]$
\begin{eqnarray}
\partial_w f_L(w)+m_b(w) f_L(w) &=& m_B f_R(w)\:,\nonumber\\
-\partial_w f_R(w)+m_b(w) f_R(w) &=& m_B f_L(w)\:.\label{12}
\end{eqnarray}
The eigenvalue, $m_B$ is the mass of the nucleon mode $B(x)$ where
4D Dirac field for the nucleon is $B=\left(\begin{array}{c} B_L \\
B_R \end{array}\right)$ and the eigenfunctions $f_{L,R}(w)$ are
normalized as
\begin{equation}
\int_{-w_{max}}^{w_{max}} dw\,\left|f_L(w)\right|^2 =
\int_{-w_{max}}^{w_{max}} dw\,\left|f_R(w)\right|^2 =1\:.\label{13}
\end{equation}
Using the properties $f_L(w)=\pm f_R(-w)$ as well as $\psi_0(w)$ and
$\psi_n(w)$ under $w\to -w$ and by plugging into the mode expansion
of gauge field, the 4D effective action is achieved for nucleon
\begin{eqnarray}
\int dx^4 (-i\bar B \gamma^\mu\partial_\mu B-im_B\bar BB+
{\cal L}_{\rm vector} +{\cal L}_{\rm axial}),\label{14}
\end{eqnarray}
where the nucleon coupling to vector mesons, ${\cal L}_{\rm
vector}$, and axial mesons, ${\cal L}_{\rm axial}$ are
\begin{eqnarray}
{\cal L}_{\rm vector}=-i\bar B \gamma^\mu \beta_\mu B-\sum_{k\ge
0}g_{V}^{(k)} \bar B \gamma^\mu  a_\mu^{(2k+1)}B,
\nonumber \\
{\cal L}_{\rm axial}=-\frac{i g_A}{2}\bar B  \gamma^\mu\gamma^5
\alpha_\mu B -\sum_{k\ge 1} g_A^{(k)} \bar B \gamma^\mu\gamma^5
a_\mu^{(2k)} B ,\label{15}
\end{eqnarray}
where various coupling constants, $g_{A,V}^{(k)}$ as well as the
pion-nucleon axial coupling, $g_A$ are calculated by suitable
wave-function overlap integrals. These coupling constants are
studied in [13] with all the specifics.
\\Finally, in general, the one boson exchange nucleon-nucleon
potential is written as [16]
\begin{equation}
V_\pi+V_{\eta'}+\sum_{k=1}^\infty V_{\rho^{(k)}}+\sum_{k=1}^\infty
V_{\omega^{(k)}} +\sum_{k=1}^\infty V_{a^{(k)}}+\sum_{k=1}^\infty
V_{f^{(k)}},\label{16}
\end{equation}
that is a sum of the pseudo-scalar, vector and axial vector mesons
exchange terms respectively.
\\But only following four classes of these couplings have a leading
contribution in nucleon-nucleon potential
\begin{eqnarray}
\frac{g_{\pi{\cal NN}}M_{KK}}{2m_{\cal N}} \sim g_{\omega^{(k)}\cal
NN} \sim \frac{\tilde g_{\rho^{(k)}\cal NN}M_{KK}}{2m_{\cal N}} \sim
g_{a^{(k)}\cal NN}.\label{17}
\end{eqnarray}
In the $D_4-D_8$ holography model, the pion mass is zero then one
pion exchange potential (OPEP) in this sense has the following form
\begin{eqnarray}
V_{\pi}=\frac{1}{4\pi}\left(\frac{g_{\pi{\cal NN}}M_{KK}}{2m_{\cal
N}}\right)^2\frac{1}{M_{KK}^2r^3}S_{12}\vec\tau_1\cdot\vec\tau_2.\label{18}
\end{eqnarray}
Also, the holographic potentials for the isospin singlet vector
mesons, $\omega^{(k)}$, isospin triplet vecto mesons, $\rho^{(k)}$
and the triplet axial-vector mesons, $a^{(k)}$ are
\begin{eqnarray}
V_{\omega^{(k)}}= \frac{1}{4\pi}\; \left(g_{\omega^{(k)}\cal
NN}\right)^2\;m_{\omega^{(k)}}\; y_0(m_{\omega^{(k)}} r),\label{19}
\end{eqnarray}
and
\begin{eqnarray}
&&V_{\rho^{(k)}}\simeq
 \frac{1}{4\pi} \left(\frac{\tilde g_{\rho^{(k)}\cal
 NN}M_{KK}}{2m_{\cal N}}\right)^2 \frac{m_{\rho^{(k)}}^3}{3M_{KK}^2}\times \nonumber\\
&&[ 2y_0(m_{\rho^{(k)}} r)\vec\sigma_1\cdot\vec \sigma_2
-y_2(m_{\rho^{(k)}} r) S_{12}(\hat
{r})]\vec\tau_1\cdot\vec\tau_2 ,\label{20}
\end{eqnarray}
\begin{eqnarray}
&&V_{a^{(k)}}\simeq \frac{1}{4\pi}\,\left({g_{a^{(k)}\cal
NN}}\right)^2 \frac{m_{a^{(k)}}}{3} \times \nonumber\\ &&[
-2y_0(m_{a^{(k)}} r)\vec\sigma_1\cdot\vec \sigma_2 +y_2(m_{a^{(k)}}
r) S_{12}(\hat {r})]\vec\tau_1\cdot\vec\tau_2  .\label{21}
\end{eqnarray}
respectively.
\\ In the above equations, level p is determined by
distance scale and
\begin{eqnarray}
S_{12}=3(\vec\sigma\cdot\hat r)(\sigma_2\cdot\hat r)
-\vec\sigma_1\cdot\vec\sigma_2,\label{22}
\end{eqnarray}
\begin{eqnarray}
y_0(x)=\frac{e^{-x}}{x},~~~~~~~
y_2(x)=\left(1+\frac{3}{x}+\frac{3}{x^2}\right)\frac{e^{-x}}{x}\,
.\label{23}
\end{eqnarray}
The massess of all mesons are of order $M_{KK}$ and
$m_{\rho^{(k)}}=m_{\omega^{(k)} } < m_{a^{(k)}}$.
\\In general, for large limited $\lambda$, in the smallest distance, $1/\sqrt{\lambda}M_{KK}$,
the one meson exchange potential is satisfied. Also, $p\simeq
\sqrt{\lambda/10}$ is an acceptable value for this potential.
\\ In the large $\lambda N_c$ limit, the coupling constants are
given by [13]
\begin{eqnarray}
&&\frac{g_{\pi \mathcal{NN}}}{2m_ \mathcal{N}} M_{KK} \simeq {8.43}
\sqrt{\frac{N_c}{\lambda}},
\nonumber\\
&&g_{\omega^{(k)} \mathcal{NN}} \simeq \sqrt{2.3^3.\pi^3} \hat{\psi}_{(2k-1)}(0) \sqrt{\frac{N_c}{\lambda}}=\xi_k \,
\sqrt{\frac{N_c}{\lambda}},
\nonumber \\
&&\frac{\tilde{g}_{\rho^{(k)}\mathcal{NN}}}{2m_\mathcal{N}}M_{KK}
\simeq \sqrt{\frac{2^2.3^2.\pi^3}{5}} \hat{\psi}_{(2k-1)}(0) \sqrt{\frac{N_c}{\lambda}}    =\zeta_k\,\sqrt{\frac{N_c}{\lambda}} ,
\nonumber \\
&&g_{a^{(k)}\mathcal{NN}} \simeq \sqrt{\frac{2^2.3^2.\pi^3}{5}} \hat{\psi'}_{(2k)}(0) \sqrt{\frac{N_c}{\lambda}} = \chi_k
\,\sqrt{\frac{N_c}{\lambda}}.\label{24}
\end{eqnarray}
where the coefficients, $\xi_k$, $\zeta_k$, $\chi_k$, are calculated
using the $\psi$ values by numerical methods. The value of these coefficients, calculated in [16], are listed in table I.
\begin{table}[htb]
\caption{\small Numerical results for $\hat\psi_{(2k-1)}(0)$,
$\hat\psi'_{(2k)}(0)$, $\xi_k$, $\zeta_k$ and $\chi_k$ for spin one
mesons interacting with nucleons[16]. }
\begin{tabular}{|c||c|c|c||c|c|}
\hline$\quad k\quad $ & $\hat\psi_{(2k-1)}(0)$ & $\quad\xi_k\quad$ &
$\quad \zeta_k \quad$
 &  $\hat\psi'_{(2k)}(0)$ & $\quad\chi_k\quad$  \\
\hline\hline
1 & 0.5973 & 24.44 & 8.925  &0.629 & 9.40 \\
\hline
2 & 0.5450 & 22.30 & 8.143  &1.10  & 16.4 \\
\hline
3 & 0.5328 & 21.81 & 7.961  &1.56  & 23.3\\
\hline
4 &0.5288 & 21.64 & 7.901  &2.02  & 30.1\\
\hline
5 & 0.5270 & 21.57 & 7.874  & 2.47 & 36.9\\
\hline
6 & 0.5261 & 21.52 & 7.860  & 2.93 & 43.8 \\
\hline
7 & 0.5255 & 21.50 & 7.852 & 3.38 & 50.5\\
\hline
8 & 0.5251 & 21.48 & 7.846  & 3.83 & 57.3\\
\hline
9  & 0.5249 & 21.48 & 7.843  & 4.29 & 64.1\\
\hline 10  & 0.5247 & 21.47 & 7.840  & 4.74 & 70.9\\
\hline
\end{tabular}
 \label{table1}
\end{table}
\section{ The Binding Energy of Deuteron and Tritium }
Here we aim to calculate the binding energy of deuteron and tritium
nuclei using the holographic nucleon-nucleon potential represented
in section 3.
\\ To calculate the binding energy of deuteron, the
following potential is considered
\begin{equation}
V^{holography}_{deuteron}=V_C+(V_T^{\sigma}\vec\sigma_1\cdot\vec\sigma_2+
V_T^{S} S_{12})\,\vec\tau_1\cdot\vec\tau_2 .\label{25}
\end{equation}
where
\begin{equation}
V_C =\sum_{k=1}^{10} \frac{1}{4\pi}\; \left(g_{\omega^{(k)}\cal
NN}\right)^2\;m_{\omega^{(k)}}\; y_0(m_{\omega^{(k)}} r)m,
\label{26}
\end{equation}

\begin{eqnarray}
V_T^{\sigma}&&=\sum_{k=1}^{10} \frac{1}{4\pi} \left(\frac{\tilde
g_{\rho^{(k)}\cal NN}M_{KK}}{2m_{\cal N}}\right)^2
\frac{m_{\rho^{(k)}}^3}{3M_{KK}^2} [ 2y_0(m_{\rho^{(k)}}
r)]\nonumber\\
&&+\sum_{k=1}^{10} \frac{1}{4\pi}\,\left({g_{a^{(k)}\cal
NN}}\right)^2 \;\frac{m_{a^{(k)}}}{3} \;[ -2y_0(m_{a^{(k)}}
r)],\label{27}
\end{eqnarray}
and,
\begin{eqnarray}
V_T^{S}&&=\frac{1}{4\pi}\left(\frac{g_{\pi{\cal NN}}M_{KK}}{2m_{\cal
N}}\right)^2\frac{1}{M_{KK}^2r^3}\; \nonumber\\
&&+\sum_{k=1}^{10} \frac{1}{4\pi} \left(\frac{\tilde
g_{\rho^{(k)}\cal NN}M_{KK}}{2m_{\cal N}}\right)^2
\frac{m_{\rho^{(k)}}^3}{3M_{KK}^2} [ -y_2(m_{\rho^{(k)}} r)] \; \nonumber\\
&&+\sum_{k=1}^{10} \frac{1}{4\pi}\,\left({g_{a^{(k)}\cal
NN}}\right)^2 \;\frac{m_{a^{(k)}}}{3} \;[ y_2(m_{a^{(k)}} r)]
.\label{28}
\end{eqnarray}
\\The values of couplings constants for different amount of k, along with
the mass of the vector and axial vector mesons (in unit of $M_{KK}$
and for large $\lambda\,N_c$) are presented in the table II. In
these computations, we choose $\lambda=400$, $ m_{\cal N}\,=\,0.55
GeV$ and $N_c=3$ for realistic QCD.

\begin{table}[htb]
\caption{\small Numerical results for masses and coupling constants
for spin one mesons interacting with nucleons in the large $\lambda
N_c $ limit. We choose $\lambda=400$, $ m_{\cal N}\,=\,0.55 GeV$ and
$N_c=3$ for realistic QCD.}
\begin{tabular}{|c||c|c||c|c|c|}
\hline  $\quad k \quad$&$\quad m_{\omega^{(k)}}$&$\quad
m_{a^{(k)}}$&$ \quad g_{\omega^{(k)}\mathcal{NN}}$&$\quad
\widetilde{g}_{\rho^{(k)}\mathcal{NN}}$& $ \quad
g_{a^{(k)}\mathcal{NN}}$
\\
\hline \hline
1  & 0.818  & 1.25 &  2.1165  &  0.7055  &  0.8140    \\
\hline
2  & 1.69   & 2.13 &  1.9312  &  0.6437  &  1.4202    \\
\hline
3  & 2.57   & 3.00 &  1.8888  &  0.6296  &  2.0178    \\
\hline
4  & 3.44   & 3.87 &  1.8740  &  0.6246  &  2.6067    \\
\hline
5  & 4.30   & 4.73 &  1.8680  &  0.6226  &  3.1956    \\
\hline
6  & 5.17   &5.59  &  1.8636  &  0.6212  &  3.7931    \\
\hline
7  & 6.03   & 6.46 &  1.8619  &  0.6206  &  4.3734    \\
\hline
8  & 6.89   & 7.32 &  1.8602  &  0.6200  &  4.9623    \\
\hline
9  & 7.75   & 8.19 &  1.8602  &  0.6200  &  5.5512    \\
\hline
10 & 8.62  &  9.05 &  1.8593  &  0.6197  &  6.1401    \\
\hline
\end{tabular}
\end{table}
The deuteron nucleus consist of one proton and one neutron, thus by
superselection rules we have
\begin{equation}
S_{12}=2,\,\,\,\,\,\,\,\vec\sigma_1\cdot\vec\sigma_2 =
1,\,\,\,\,\,\,\, \vec\tau_1\cdot\vec\tau_2=-3.\label{29}
\end{equation}
The deuteron potential for the large $N_c$ with $p=10$ is calculated
and has been shown in figure 1. As it is clear from this figure, the
deuteron potential has a minimum point at $ 4.41 \, M_{KK}$. For
distance, r less than $r_{min}$ potential increases rapidly and
become infinity at $r=0$ as expected. The minimum value of potential
is $ -1.9645\, M_{KK}\, N_c/ 4\pi \lambda $. So the binding energy
of deuteron is obtained roughly $-2.204 \, MeV $ that is consistent
with the experimental nuclear data.
\begin{figure}[bth]
\centerline{\includegraphics[width=10cm]{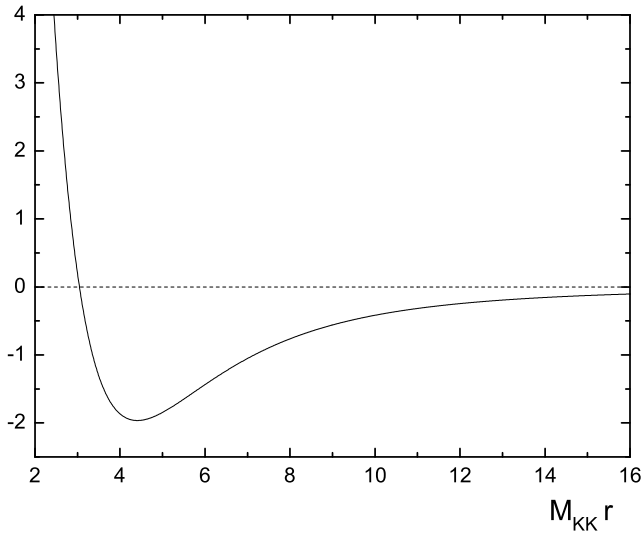}}
\caption{\small The deuteron potential in large $\lambda\, N_c$
limit and p=10. The horizontal axis is $r\,M_{KK}$, while the
potential energy along the vertical axis is in unit of $M_{KK}\,
N_c/ 4\pi \lambda $. \label{fig1}}
\end{figure}
\\Also, tritium consist of three nucleons, two neutrons and one proton,
so we suppose the following form for the tritium potential
\begin{eqnarray}
V^{holography}_{Tritium}&=& V_{12}+ V_{13}+ V_{23} \nonumber\\
&=&V_C+(V_T^{\sigma}\vec\sigma_1\cdot\vec\sigma_2+ V_T^{S}
S_{12})\,\vec\tau_1\cdot\vec\tau_2 \nonumber\\
&+&V_C+(V_T^{\sigma}\vec\sigma_1\cdot\vec\sigma_3+ V_T^{S}
S_{13})\,\vec\tau_1\cdot\vec\tau_3 \nonumber\\
&+&V_C+(V_T^{\sigma}\vec\sigma_2\cdot\vec\sigma_3+ V_T^{S}
S_{23})\,\vec\tau_2\cdot\vec\tau_3.\label{30}
\end{eqnarray}
The superselection rules for this three-nucleon systems implies that
\begin{eqnarray}
&&S_{12}=2\,\,,\,\,\,\,\,\vec\sigma_1\cdot\vec\sigma_2 =1
\,\,,\,\,\,\,\,\,\,\,\,\, \vec\tau_1\cdot\vec\tau_2=-3 \nonumber\\
&&S_{13}=0\,\,,\,\,\,\,\,\vec\sigma_1\cdot\vec\sigma_3 =-3
\,\,,\,\,\,\,\, \vec\tau_1\cdot\vec\tau_3= -3\nonumber\\
&&S_{23}=0\,\,,\,\,\,\,\,\vec\sigma_2\cdot\vec\sigma_3 =-3
\,,\,\,\,\,\,\, \vec\tau_2\cdot\vec\tau_3=1 \, \, .\label{31}
\end{eqnarray}
The holographic potential of tritium in terms of  $M_{KK}\,r$ has
been shown in figure 2. This potential also has a minimum that
occurs in $ 7.46\, M_{KK}$. The value of potential in its minimum is
$ -0.617\, M_{KK}\, N_c/ 4\pi \lambda $, so the binding energy of
tritium is equal to $-1.034\, MeV$. This figure also shows the
repulsive behavior at short distances.
\begin{figure}[bth]
\centerline{\includegraphics[width=10cm]{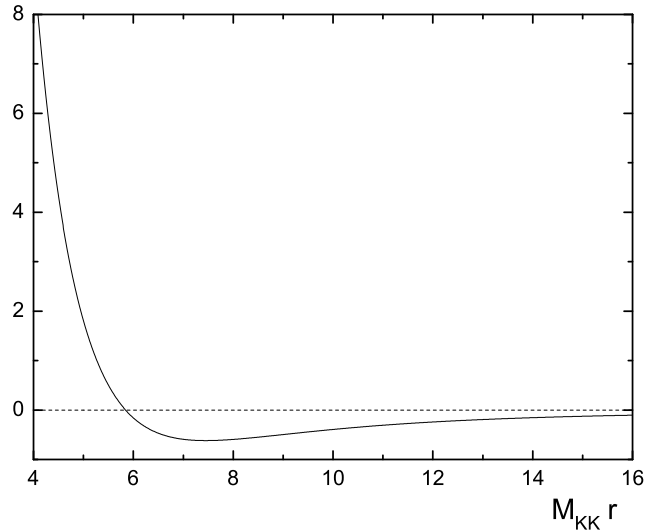}}
\caption{\small The Tritium potential in large $\lambda\, N_c$ limit
and p=10. The horizontal axis is $r\,M_{KK}$, while the potential
energy along the vertical axis is in unit of $M_{KK}\, N_c/ 4\pi
\lambda $. \label{fig2}}
\end{figure}
\\
\section{Conclusion}
In this investigation we calculated the deuteron and tritium binding
energy using the QCD holography model. Here we used the
nucleon-nucleon interaction in the $D_4-D_8$ model in the base of
one-boson exchange picture. This potential involves only the
exchanges of pions, isospin singlet mesons, isospin triplet mesons
and triplet axial-vector mesons. We selected the $\lambda=400$ and
at least 10 terms of infinite tower of spin one mesons are
considered.
\\ We depicted the deuteron and tritium potential in terms of $M_{KK}
r$ and in unit of $M_{KK}N_c/4\pi\lambda$. As it is indicated in
figures 1 and 2, these potentials have repulsive behavior at short
distances and became roughly zero at large $M_{KK} r$. The deuteron
potential have a shallow minimum in depth $\sim -13.84 \,M_{KK}\,N_c
/ \lambda$ around the $ r\,M_{KK}= 4.41  $. The tritium potential
too, have a more shallow minimum around the $r \,M_{KK} = 7.46 $
with depth $\sim \,-4.35\, M_{KK}\,N_c/ \lambda$. Thus by using this
information the binding energies of deuteron and tritium nuclei
approximated by $-2.204\, MeV$ and $ -1.039 \,MeV$, respectively.
\\This method may be improved to obtain binding energies of heavier
nuclei by considering exchange of heavier mesons.

\end{document}